\documentclass[preprint2]{aastex631}

\usepackage{xspace}
\usepackage{amsmath}

\newcommand{\funits}{erg~cm$^{-2}$~s$^{-1}$}

\newcommand{\nustar}{\textit{NuSTAR}\xspace}

\newcommand{\swift}{\textit{Swift}\xspace}
\newcommand{\chandra}{\textit{Chandra}\xspace}

\newcommand{\lunits}{erg~s$^{-1}$}
\newcommand{\punits}{ph~cm$^{-2}$~s$^{-1}$}

\received{June 7, 2023}

\submitjournal{ApJL}

\begin{document}

\title{Early hard X-rays from the nearby core-collapse supernova SN2023ixf}

\correspondingauthor{Brian Grefenstette}
\email{bwgref@srl.caltech.edu}

\author[0000-0002-1984-2932]{Brian W. Grefenstette}
\affiliation{Space Radiation Laboratory
California Institute of Technology 
1200 E California Blvd 
Pasadena, CA 91125, USA}

\author[0000-0002-8147-2602]{Murray Brightman}
\affiliation{Space Radiation Laboratory
California Institute of Technology 
1200 E California Blvd 
Pasadena, CA 91125, USA}

\author[0000-0001-5857-5622]{Hannah P. Earnshaw}
\affiliation{Space Radiation Laboratory
California Institute of Technology 
1200 E California Blvd 
Pasadena, CA 91125, USA}

\author{Fiona A. Harrison}
\affiliation{Space Radiation Laboratory
California Institute of Technology 
1200 E California Blvd 
Pasadena, CA 91125, USA}

\author[0000-0003-4768-7586]{R.~Margutti}
\affil{Department of Astronomy, University of California, Berkeley, CA 94720, USA}
\affil{Department of Physics, University of California, Berkeley, CA 94720, USA}

\begin{abstract}
We present \nustar  observations of the  nearby SN\,2023ixf in M101 ($d=6.9$\,Mpc) which provide the earliest hard X-ray detection of a non-relativistic stellar explosion to date at $\delta t\approx$\,4\,d and $\delta t\approx$\,11\,d. The spectra are well described by a hot thermal bremsstrahlung continuum with $T>25\,\rm{keV}$ shining through a thick neutral medium with a neutral hydrogen column that decreases with time (initial $N_{\rm{Hint}}=2.6\times 10^{23}\,\rm{cm^{-2}}$). A prominent neutral Fe K$\alpha$ emission line is clearly detected, similar to other strongly interacting SNe such as SN2020jl. The rapidly decreasing intrinsic absorption with time suggests the presence of a dense but confined circumstellar medium (CSM). The absorbed broadband X-ray luminosity (0.3--79 keV) is $L_{X} \approx 2.5 \times 10^{40}$ \lunits\ during both epochs, with the increase in overall X-ray flux related to the decrease in the absorbing column. Interpreting these observations in the context of thermal bremsstrahlung radiation originating from the interaction of the SN shock with a dense medium we infer large particle densities in excess of $n_{\rm{CSM}}\approx 4\times  10^{8}\,\rm{cm^{-3}}$ at $r<10^{15}\,\rm{cm}$, corresponding to an enhanced progenitor mass-loss rate of $\dot M \approx 3\times 10^{-4}$ M$_{\odot}$\,yr$^{-1}$ for an assumed wind velocity of $v_w=50$\,km\,s$^{-1}$.

\end{abstract}



\section{Introduction} \label{sec:intro}
Observations of stellar explosions in the past decade have revealed the complex mass-loss history of massive stars in the centuries leading up to core collapse. The observational evidence has been accumulating from a variety of independent channels, including the direct detection of stellar outbursts before the stellar explosion, the presence of narrow spectral lines originating from the ionization of material ahead of the explosion's shock, as well as luminous UV, X-ray and radio emission (e.g., \citealt{Pastorello07,Pastorello13,Pastorello18,Margutti14,Ofek14,Strotjohann21,Jacobson-Galan22, Morozova18,Morozova20,Bostroem19,Dessart22,Schlegel90,Filippenko97,Pastorello08,Perley22,Chevalier06,Soderberg062003bg,Dwarkadas10,Chevalier17,Stroh21}). The consequent inference of highly time-dependent mass loss in evolved massive stars challenges our current understanding of massive star evolution and points at a new evolutionary path where nuclear burning instabilities and interaction with a binary companion play primary roles (e.g., \citealt{Quataert12,Smith14,Smith14b}). 

The SN shock interaction with the circumstellar medium  (CSM) is a well-known source of copious X-ray and radio emission (e.g., \citealt{Chevalier17}). For young SNe exploding in dense media, the X-ray spectrum is expected to extend to the hard X-rays and to be dominated by thermal bremsstrahlung emission with high characteristic temperatures of $T\gtrsim 10^{7}\,\rm{K}$ that result from fast shock velocities; $v_{sh}\gtrsim 10^4\,\rm{km\,s^{-1}}$ (e.g. \citealt{Fransson96}), as was confirmed by observations of SNe 2010jl and 2014C \citep{Margutti2017, Ofek14}. However, the hard X-ray part of a SN spectrum has been very rarely sampled so far, and only four SNe have been detected (all with \nustar): the H-poor, type-Ib SN\,2014C  \citep{Margutti2017,Brethauer22,Thomas22}, the H-rich type-IIn SN2010jl \citep{Ofek14,Chandra15}, the type-IIP SN2017eaw \citep{Grefensetette17}, and the remarkable fast blue optical transient AT\,2018cow \citep{Margutti2019}. Here we present \nustar observations of the closest core-collapse supernova in the last decade, SN2023ixf, with the earliest detection of a non-relativistic SN in the hard X-rays.

SN2023ixf is a Type II supernova discovered by Koichi Itagaki in M101 (NGC 5457; \citealt{2023TNSTR1158....1I}). The discovery date was 2023-05-19 17:27:15 UTC, although serendipitous pre-discovery data by Chinese amateur astronomers indicate the onset of the supernova at 2023-05-18T20:30 UTC \citep{2023TNSAN.130....1M}, so we adopt the latter as $T_{0}$. We use a distance to M101 of 6.90 Mpc \citep{Riess22}. The observed optical spectrum showed significant evolution, with the early spectra displaying prominent flash ionization spectral features  \citep{2023TNSAN.119....1P} that subside by $\approx$10-d after the explosion.

In this Letter we describe the early \nustar observations along with \swift monitoring, provide an overview of our results, and discuss the implications of the hard X-ray emission and evolution.

\section{Observations}

After the identification of SN2023ixf as a nearby Type II supernova, we requested a Director's Discretionary Time (DDT) Target of Opportunity (ToO) observation using \nustar, \citep[the \textit{Nuclear Spectroscopic Telescope ARray,}][]{2013ApJ...770..103H}. \nustar began observing at 2023-05-22T17:56:09 ($\approx3.9\,$d) with a total exposure time of 46-ks, representing just over a day of elapsed time. A second epoch of DDT time was requested (starting at $\approx10.5\,$d) following the first \nustar detection. Details of the \nustar observations are given in Table \ref{tab:obs}.

\startlongtable
\begin{deluxetable*}{c|cc|c|c}
\tablecaption{\nustar Observations \label{tab:obs}}
\tablehead{
 \colhead{Sequence ID} & \colhead{Obs Start} & \colhead{Obs End} & \colhead{Exposure (ks)} & \colhead{$\delta$t$^{a}$ (days)}
}
\startdata
90302004002 & 2023-05-22T17:56:09 & 2023-05-23T16:11:09 & 42 & 3.9-4.8 \\
90302004004 & 2023-05-29T08:56:09 & 2023-05-30T07:06:09 & 42 & 10.5-11.4 \\
\enddata
\tablecomments{$^{a}$Age is given with respect to the onset time of 2023-05-18T20:30:00 UTC}
\label{tab:obs}
\end{deluxetable*}

\begin{figure*}[ht!]
\centering
\includegraphics[height=8cm]{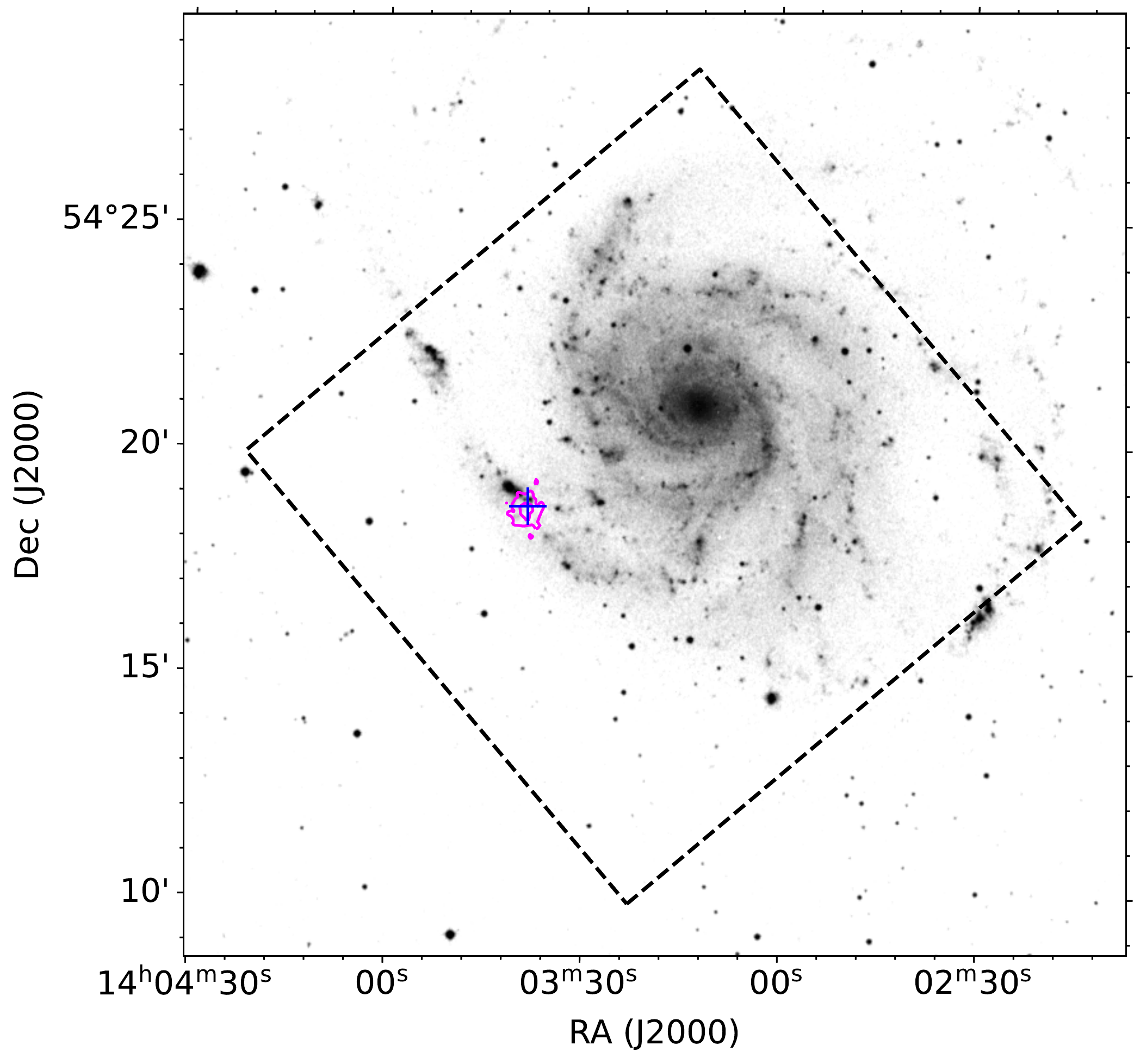}
\includegraphics[height=8cm]{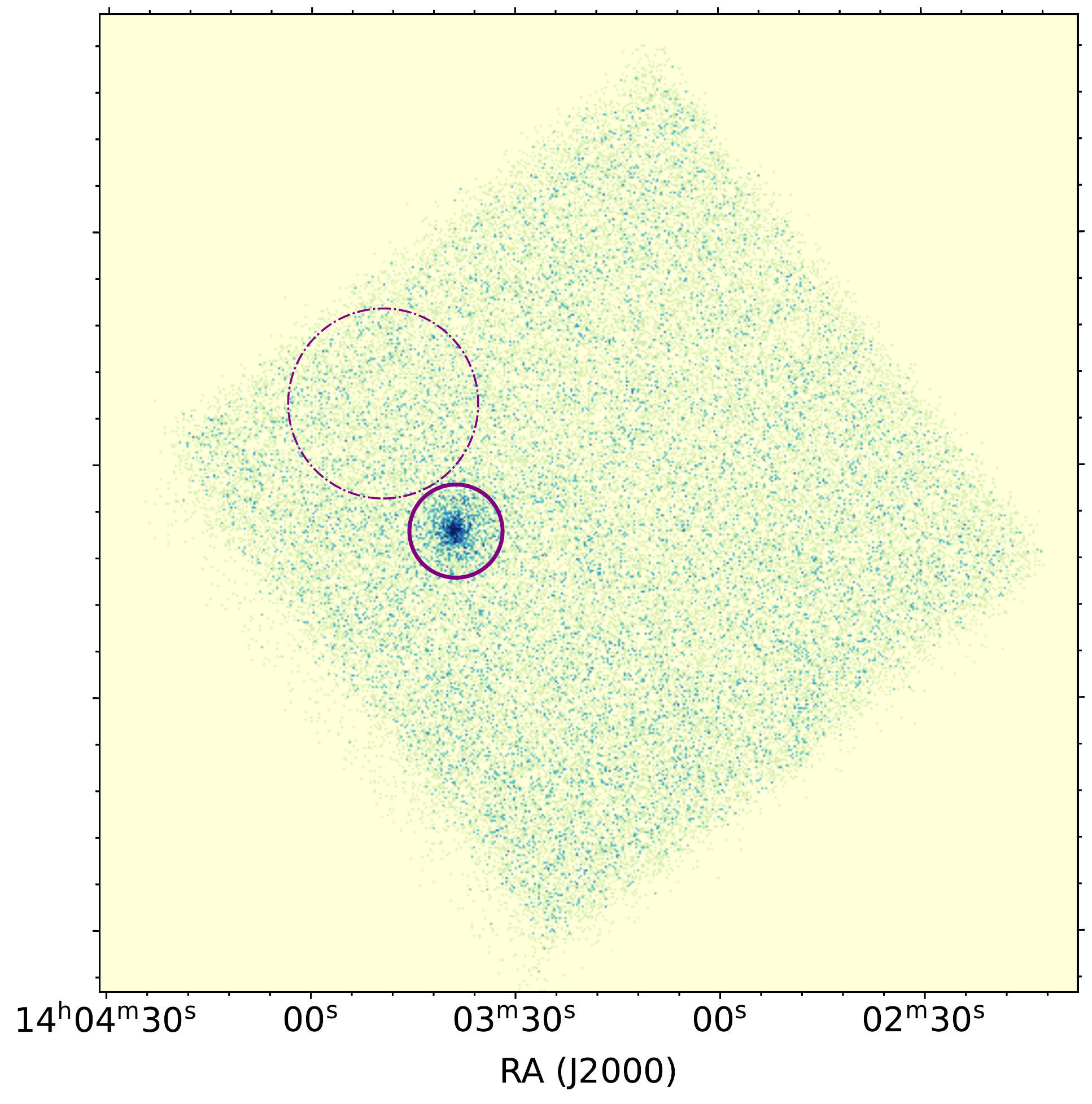}
\caption{\textit{Left}, Digitized Sky Survey image of M101, with the \nustar field of view indicated with a black dashed line, the \nustar contours indicated in magenta, and the optical measured position of SN2023ixf indicated with a blue cross. \textit{Right}, the \nustar image, with the source extraction region marked with a solid purple circle, and the nearby background region marked with a dot-dashed circle.  
\label{fig:fig1}}
\end{figure*}

We produced high level science products (spectra and response files) using the standard \textsc{nustardas} analysis tools with an extraction region of radius of 60\arcsec. A nearby region (Fig.~\ref{fig:fig1}) was used to estimate the local backgrounds. Two X-ray sources, NGC 5457 X-19 and NGC 5457 X-20, are within the \nustar\ extraction region, however, upon examining these two sources using the latest archival \chandra\ observation of the galaxy (observation 19304 taken in 2017), they are both soft and contribute $<1\times 10^{-14}$ \funits~in the 3--10\,keV energy band, so we do not expect them to make a significant contribution to the \nustar\ spectrum.

The Neil Gehrels Swift Observatory \citep{2004ApJ...611.1005G} also observed the SN 25 times from 2023-05-20T06:24:57 to 2023-05-25T15:03:54 (sequence IDs 00016038001--12, 00016043001--2 and 00032481002--17). We used the online tool provided by the University of Leicester\footnote{https://www.swift.ac.uk/user\_objects/} \citep{2007A&A...469..379E,2009MNRAS.397.1177E} to carry out source detection, and generate images, the lightcurve and spectrum. We do not allow centroiding of the data since this is known to fail for faint sources such as this.

\section{Results}

\subsection{NuSTAR Spectroscopy}

In both \nustar\ telescopes, a point source is clearly detected up to $\approx$30 keV in both epochs. This is the only source easily identified in M101 (Fig.~\ref{fig:fig1}). The two co-aligned telescopes detect the source and report  offsets of 6\arcsec\ and 3\arcsec\ from the optical position of the supernova explosion (14:03:38.562, +54:18:41.94, \citealt{2023TNSTR1158....1I}). This is consistent with the SN location given the systematic errors in the \nustar absolute astrometry.

There is a strong evolution of the source between Epoch I and Epoch II, demonstrating that the \nustar emission is clearly dominated by the SN. We simultaneously fit the un-binned data from both telescopes using the W-statistic. Data are re-binned for plotting purposes below.

\begin{figure*}[ht!]
\plotone{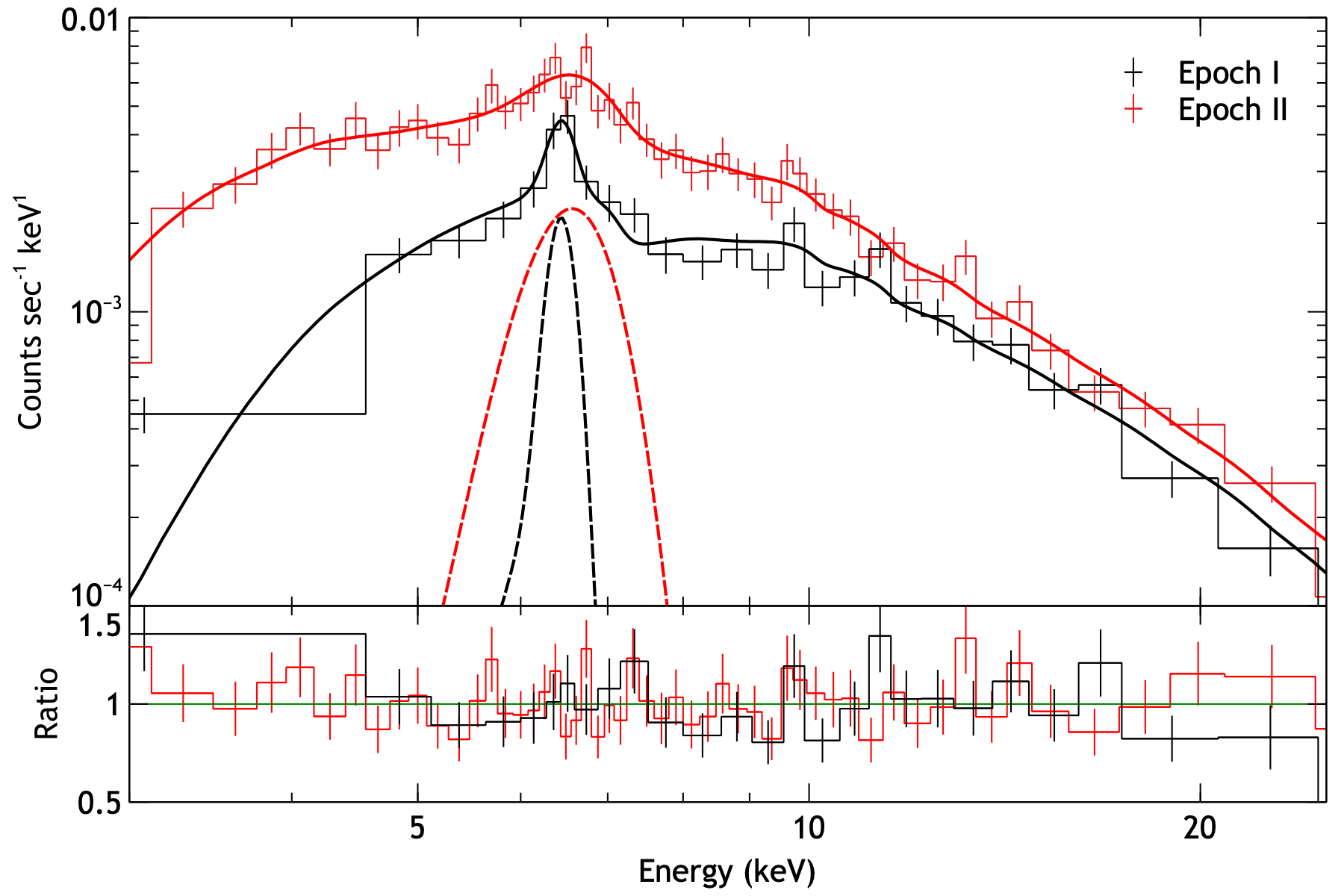}
\caption{(\textit{Top}) The background-subtracted spectrum for Epoch I ($\delta$t $\approx\,$4-d, black) and Epoch II ($\delta$t $\approx\,$11-d, red) showing the best-fit model (solid lines) and the Gaussian component (dotted lines) demonstrating the spectral evolution of the source. (\textit{Bottom}) Residuals to the best-fit model.
\label{fig:speccounts}}
\end{figure*}

In Epoch I, the spectra for both telescopes show a clear turnover to low energies (e.g., a high absorption column) as well as a significant excess near the Fe line region around 6.4 keV. The width of the line in Epoch I is consistent with the energy resolution of the \nustar detectors, so we can only report an upper limit on the line width. In Epoch II, the overall source flux has increased below 10 keV, and the Fe line features have become slightly broader and shifted to higher energies (Fig.~\ref{fig:speccounts}). The change in the Fe line centroid and width may also be a blending of emission from various lines at higher ionization energies that are not resolved by \nustar.

In both epochs we adopt a spectral model that consists of an underlying bremsstrahlung spectrum with neutral absorption along the line of sight. We include an additive line component to account for reprocessing of the high energy emission in the absorbing material and/or the presence of ionized lines in the data. For all models we use the solar abundances of \cite{1989GeCoA..53..197A}. In \textsc{XSPEC}, our model is \texttt{tbabs(nlapec+gauss)}. Galactic absorption along this line of sight is low enough\footnote{https://heasarc.gsfc.nasa.gov/cgi-bin/Tools/w3nh/w3nh.pl} (8 $\times 10^{20}$ cm$^{-2}$) that we neglect it for \nustar analysis. Below we adopt $N_{\rm{Hint}}$ to refer to the intrinsic absorbing material in the SN.

To estimate the uncertainties on the fit parameters we use the Markov Chain Monte Carlo (MCMC) method \citep[\texttt{emcee,}][]{foreman-mackey_emcee:_2013} as implemented in {\sc XSPEC} to estimate 90\% confidence intervals for all parameters. Table \ref{tab:spec_fits} provides the best-fit parameters and their confidence intervals.

\subsection{Swift-XRT observations}
For the \swift data, while the supernova is not detected in the full 0.3--10 keV band in a stack of the first 25 observations which span 6 days with a total exposure time of 36.6 ks, it is detected in the 2--10 keV band with a count rate of $7.7^{+2.4}_{-2.1}\times10^{-4}$ count s$^{-1}$. We fit the stacked spectrum with our fiducial model fixing all parameters to the \nustar\ Epoch I values, using a multiplicative constant to allow for flux variability. We find an observed 0.3--10 keV flux of $6.6^{+10}_{-6.6}\times10^{-14}$ \funits, $\approx$9$\times$ lower than measured by \nustar\ during Epoch I, implying some X-ray flux evolution. If we limit our analysis to the 6 observations that occurred during the \nustar\ Epoch I observation, the observed 0.3--10 keV flux is $3.4^{+2.9}_{-2.5}\times10^{-13}$ \funits, which is consistent with the \nustar\ Epoch I flux extrapolated into this band. Unfortunately there were no \swift-XRT\ observations that took place during \nustar\ Epoch II so we cannot repeat our analysis for that observation.

No source is listed at the position of the supernova in the \chandra\ Source Catalog \citep[CSC2,][]{evans10}, and the sensitivity of the \chandra\ observations at the position of the supernova is listed by CSC2 as $8\times10^{-16}$ \funits\ in the 0.5--8 keV band and $1.2\times10^{-15}$ \funits\ in the 2--8 keV band, well below the \nustar\ and \swift-XRT fluxes. We examined the deepest archival \chandra\ image of the region, observation 934 taken in 2000 for an exposure of 98\,ks, and found the background flux within 20 arcsec of the supernova location to be $\sim1.5\times10^{-14}$ \funits\ in the 0.3--10\,keV energy range. We are therefore confident that the supernova emission dominates the \swift-XRT\ flux during these observations.

\begin{deluxetable*}{l|ccccccccc}
\tablecaption{Spectral Fits for \texttt{tbabs(nlapec+gauss)} \label{tab:spec_fits}}
\tablehead{
\colhead{Epoch} &
$N_{\rm{Hint}}^{a}$ & \colhead{$kT$ (keV)} &
\colhead{norm$^{b}$} & \colhead{Line (keV)} & 
 \colhead{Width (keV)} &
 \colhead{norm$^{c}$} &
\colhead{{\tt Wstat} / dof}
}
\startdata
Epoch I &
26${_{-7}^{+5}}$ & $>$25 & 1.06 $\pm$ 0.13 & 6.45 $\pm$ 0.08 & $<$0.2 & 6.6 $\pm$ 2 & 888 / 843  \\
Epoch II &
5.6 $\pm$ 2.7 & 34$^{+22}_{-12}$ &
1.3$^{+0.2}_{-0.1}$ &
6.57 $\pm$ 0.17 & 
0.45 $\pm$ 0.2 & 14 $\pm$ 5 & 
892 / 842 
\enddata
\tablecomments{Uncertainties indicate the 90\% confidence intervals based on the MCMC run. \\ $^{a}$10$^{22}$ atoms cm$^{-2}$; $^{b}$\texttt{nlapec} normalization  [10$^{-3}$] ; $^{c}$10$^{-6}$ \punits ; *frozen since the line is narrower than the energy resolution of the \nustar detectors. }

\end{deluxetable*}

\begin{deluxetable*}{l|c|c|c|c|c}
\tablecaption{Computed values \label{tab:spec_flux}}
\tablehead{
\colhead{Epoch} &
\colhead{EM$^{a}$} & 
\colhead{Flux$^{b}_{0.3-10~\rm keV}$} &
\colhead{Lum$^{c}_{0.3-10~\rm keV}$} &
\colhead{Flux$^{b}_{10-79~\rm keV}$} &
\colhead{Lum$^{c}_{10-79~\rm keV}$}
}
\startdata
Epoch I &
6.0 $\pm$ 0.7 $\times 10^{62}$ &
5.9 $\pm$ 0.3 $\times 10^{-13}$ & 
0.34 $\pm$ 0.02 $\times 10^{40}$ & 
3.4$_{-1.3}^{+0.2} \times 10^{-12}$ &
1.9$_{-0.7}^{+0.1}\times 10^{40}$ \\
Epoch I$^{d}$ & - &
1.7$^{+0.7}_{-0.1} \times 10^{-12}$ &
1.0$^{+0.34}_{-0.05} \times 10^{40}$&
3.5$^{+0.3}_{-1} \times 10^{-12}$ &
2.0$^{+0.2}_{-0.7} \times 10^{40}$ \\
\hline
Epoch II &
7.5$^{+0.9}_{-0.5} \times 10^{62}$ &
1.44 $\pm$ 0.08 $\times 10^{-12}$ &
0.82 $\pm$ 0.05 $\times 10^{40}$ &
3.5 $\pm~0.9 \times 10^{-12}$ &
2 $\pm~0.5\times 10^{40}$ \\
Epoch II$^{d}$ &
- &
2.5 $\pm$ 0.3 $\times 10^{-12}$ &
1.4 $\pm$ 0.2 $\times 10^{40}$ &
3.5 $\pm~0.9 \times 10^{-12}$ &
2 $\pm~0.5\times 10^{40}$
\enddata
\tablecomments{Uncertainties indicate the 90\% confidence intervals based on the MCMC run. Distance is assumed to be 6.9 Mpc. $^{a} cm^{-3}$ ; $^{b}$ \funits; $^{c}$ \lunits. $^{d}$ Deabsorbed values.}

\end{deluxetable*}

\section{Discussion}

\subsection{Evolution of the supernova emission}

The early-time X-ray flux from SN2023ixf is typical of other Type II supernovae. However, the large absorption column in Epoch I makes it difficult to compare with other supernovae that are relatively unabsorbed. To account for this, we compute an unabsorbed flux by changing the model definition to \texttt{tbabs(cflux*nlapec + gauss)}. The \texttt{cflux} component measures the intrinsic flux in the underlying continuum. Table \ref{tab:spec_flux} provides the resulting flux in the 0.3--10 and 10--79 keV bands. For a distance of 6.9\,Mpc, this results in a large a intrinsic 0.3 - 79 keV luminosity of $\sim$10$^{40}$\,\lunits in both epochs.

The primary difference between the two \nustar epochs is the dramatic reduction in the absorbing column. The intrinsic spectrum does not appear to vary much between Epoch I and Epoch II, with the luminosity in the hard (10--79 keV) band staying effectively constant. The large uncertainties in the soft band luminosity due to the poorly-constrained $N_{\rm Hint}$ are consistent with the X-ray source emerging from behind absorbing material.

\subsection{Forward shock velocity and plasma temperature}

We assume that the material producing the X-rays has been heated by the shock from the supernova explosion. The temperature of the emission can be used to infer the velocity of the shock. Using the formalism of \cite{1996ApJ...461..993F} as in \cite{2022ApJ...939..105B}:

\begin{equation}
    T \approx 2.27 \times 10^{9} \mu v^{2}_{4}~\rm K
\end{equation}

where $\mu$ is the mean molecular weight of the shocked medium (here assumed to be 0.61 for solar-like, ionized material with equipartition between electrons and ions). For Epoch I, the \nustar spectra can only place a lower limit on the electron temperature due to the limited signal-to-noise at high energies. In Epoch II, the \nustar spectra can constrain the temperature to be $\approx$35 keV, which corresponds to a velocity of $\sim$5,400 km s$^{-1}$. We take this as an order of magnitude estimate for the actual shock velocity,  consistent with other supernova shocks.

However, we have explicitly assumed that the electrons and ions reach equipartition, which may not be correct. We measure the electron temperature $T_e$ from the spectra. The timescale for energy transfer from ions to electrons is as follows \citep[from Eq. 26 of][]{Chevalier_2006}:
\begin{equation}
\begin{split}
t_{e-i} & = \frac{4.2\times 10^{-22}}{\ln(\Lambda)}\eta(Z)\frac{T_e^{3/2}}{\rho} \\
& \approx 0.8\,\rm{d} \Big (\frac{T_e}{25\,\rm{keV}}\Big)^{3/2} \Big( \frac{9.8\times 10^{-16}}{\rm{g/cm^3}}\Big)
\end{split}
\end{equation}
where $\eta=1$ for H and $\eta=4/Z$ for heavier elements of charge $Z$; $\ln(\Lambda)\approx 30$ is the Coulomb logarithm, and we have normalized the equation to the values that apply to our first \nustar epoch.

Based on this result, complete electron-ion (e-i) equipartition is unlikely even at the time of our first \nustar epoch, as $t_{e-i}$ is comparable to the time of our first \nustar epoch. For the second \nustar epoch at $\delta t\approx11$\,d the density is lower and we derive $t_{e-i}\approx 8\,\rm{d}$: complete e-i equipartition is questionable. We can reverse this argument and calculate the minimum electron temperature at a particular time and density and compare this value to our constraint. Doing so we obtain a minimum electron temperature of $\approx$60 keV (35 keV) at the time of our first (second) \nustar epoch. Our high electron temperatures are therefore consistent with typical supernova shock velocities of $\sim10^{4}$ km s$^{-1}$ \citep{Fransson96}.

\subsection{Origin of the early Fe emission and the density of the CSM}

In the first epoch, the Fe line appears to be related to neutral (cold) Fe emission, rather than from shock-heated plasma. This is consistent with reprocessing of the X-ray emission in cold, circumstellar material responsible for the high absorption column. This is similar to the early neutral Fe lines observed in SN2010jl \citep{2012ApJ...750L...2C} which was associated with a clumpy circumstellar material. 

To test this, we also tested a model with a power-law representation of the intrinsic spectrum absorbed by a neutral, spherically distributed medium. This model \citep{brightman11} includes the re-emission of the neutral lines self-consistently as well as the effects of Compton scattering in the surrounding medium. We find that the first epoch spectrum can be reasonably fit with the same $N_{\rm Hint}$ as in the baseline model.

We can use these measurements of the absorbing material to estimate the pre-supernova mass loss rate for the star. Assuming a shock velocity of 15,000 km s$^{-1}$ places the forward shock at $R_{1}\sim5.7\times 10^{14}$ cm 4.4-d after the explosion and at $R_{2}\approx 1.4 \times 10^{15}$ cm at 11-d.  The fast disappearance of the the flash ionization spectral features \citep{2023TNSAN.145....1S, 2023arXiv230600263Y} supports a dense, but confined, CSM. All of the absorbing material is assumed to be local to the SN environment, an assumption supported by the rapid decrease in $N_{\rm Hint}$ between the two \nustar epochs. If all of the material is local to the SN explosion, then measured $N_{\rm Hint}$ at $\delta$t$\approx$ 4.4-d implies $\dot M \approx 2.5\times 10^{-4}$ M$_{\odot}$~yr$^{-1}$ for an assumed $v_w=50$\,km~s$^{-1}$, Hydrogen-dominated chemical composition, and a wind-like density profile. 

An alternative estimate of the mass local to the SN environment comes from the observed decrease of neutral hydrogen absorption ($\Delta \rm{N_{H}}$) between the two \nustar epochs  at 4.4 and 11.0\,d. This assumes that the measured $\Delta \rm{N_{H}}$ is largely a consequence of the shock plowing through the medium and emerging from the dense circumstellar material or that newly ionized material was mostly located between $R_{1}$ and $R_{2}$. The inferred mass-loss rate (for H-dominated composition and $v_w=50$\,km/s) is $\dot M \approx 3.0\times  10^{-4}$ M$_{\odot}$~yr$^{-1}$. For these parameters, the amount of circumstellar mass sampled by the shock during the first $\sim11$\,d is $M_{\rm{CSM}}\approx 1.7\times 10^{-3}\,\rm{M_{\odot}}$. The inferred particle density at $R_{1}$ is $\sim 6\times 10^8\,\rm{cm^{-3}}$, decreasing to $\sim 3\times 10^7\,\rm{cm^{-3}}$ at $R_{2}$.

We can also estimate of the density of the emitting region from the Emission Measure, $EM \equiv \int n_e n_i\,dV$, where the integral is over the X-ray emitting region, and the derived density is that of shock compressed material. For a strong shock applicable here we expect this density to be four times the unshocked CSM density (which is what we measured from the $\Delta \rm{N_{H}}$ above). From our broad-band X-ray modeling we infer $EM(4.4\,\rm{d})\approx 5.7\times 10^{62}\,\rm{cm^{-3}}$ and $EM(11.0\,\rm{d})\approx 7.6\times 10^{62}\,\rm{cm^{-3}}$. Assuming a Hydrogen composition, complete ionization (consistent with the large T measured), constant particle density in the X-ray emitting region, spherical shell-like geometry with $\Delta R\approx 0.1 R$, the inferred pre-shock particle density at $R_{1}$ is $\sim 4\times 10^8\,\rm{cm^{-3}}$, decreasing to $\sim  10^8\,\rm{cm^{-3}}$ at $R_{2}$. This density estimate is consistent with the inferences from the $\Delta \rm{N_{H}}$ to within a factor of a few and thus supports the conclusion that the emitting region is the shocked CSM.

\subsection{Consistency with early radio non-detections}

Radio non-detections of SN\,2023ixf have been reported by \cite{Berger23}, \cite{matthews23} and \cite{chandra23}. The flux limits reported are as follows. SMA: $F_{\nu}<1.5\,\rm{mJy}$  (3 RMS) $\delta t \approx 3\,$d $\nu=230$\,GHz (2023 May 21.17 UT, \citealt{Berger23}). GMRT: $F_{\nu}<75\,\rm{\mu Jy}$ (3 RMS) at $\nu=1.255$\,GHz $\delta t \approx 4\,$d (2023 May 22.74 UT, \citealt{chandra23}). VLA: $F_{\nu}<33\,\rm{\mu Jy}$ (5 RMS) $\nu=10$\,GHz $\delta t \approx 4\,$d (May 23rd 2023 beginning on-source at 00:24:09 UT, \citealt{matthews23}). 

If we assume the densities derived from the $\Delta \rm{N_{H}}$ above, the (external) free-free optical depth $\tau_{\rm{230\,GHz}}(3\,d)\approx 200 \Big ( \frac{T_e}{10^4\,K} \Big )^{-1.35}$, decreasing to below 1 at $\sim 20$\,d; $\tau_{\rm{10\,GHz}}(4\,d)\approx 7\times 10^4$ decreasing to 1 at 150 days. Here the temperature $T_e$ is the temperature of the electrons in the unshocked region (while the X-ray emitting electrons are in the shocked region and at a higher temperature). The 230 GHz SMA non-detection and the 10 GHz VLA non-detection are thus not surprising and consistent with the densities inferred from the X-ray modeling.

\section{Conclusions} \label{Sec:conc}
We report on the first hard X-ray detections of SN\,2023ixf, obtained by \nustar at roughly four and eleven days after the onset of the supernova explosion. The early X-ray spectrum is highly absorbed with neutral Fe line emission produced through reprocessing of high energy photons in the circumstellar material. The absorbing column dropped substantially between the two \nustar epochs, indicating that the absorbing material is local to the supernova explosion. We infer that the CSM is confined to near the SN through the rapid disappearance of ``flash ionization" optical spectral features by $\approx$10-d combined with the rapid decrease in the $N_{Hint}$ between the \nustar epochs.

We conclude that the X-ray emission originates from the forward shock region as it interacts with the CSM. We estimate the mass-loss rate and density of the CSM to be a few 10$^{-4}$\,M$_{\odot}$\,yr$^{-1}$; for typical forward shock and wind velocities, we note that at these early times the forward shock is interacting with CSM that is only a few to 10 years old. The densities inferred for the above mass loss imply optical depths high enough to suppress the radio emission at these early times, which is consistent with the current non-detection of the supernova in the radio. If the density profile that we infer at small radii extends to larger radii, then this SN may become radio-mm bright in the next few weeks to months, while a truncated density profile would lead to an earlier emergence of the radio-mm signal.

\section*{Acknowledgments}
This work makes use of data from the \nustar mission, a project led by the California Institute of Technology, managed by the Jet Propulsion Laboratory, and funded by the National Aeronautics and Space Administration. We thank the \nustar Operations, Software and Calibration teams for support with the execution and analysis of these observations. This research has made use of the \nustar Data Analysis Software (NuSTARDAS) jointly developed by the ASI Science Data Center (ASDC, Italy) and the California Institute of Technology (USA). This research has made use of data and/or software provided by the High Energy Astrophysics Science Archive Research Center (HEASARC), which is a service of the Astrophysics Science Division at NASA/GSFC.

BG, MB, and HE acknowledge support under NASA Contract No. NNG08FD60C. R.M.\ acknowledges support by the National Science Foundation under Award No. AST-2221789 and AST-2224255.

%

\vspace{5mm}
\facilities{NuSTAR, Swift(XRT)}


\software{astropy \citep{2013A&A...558A..33A,2018AJ....156..123A,2022ApJ...935..167A},  
          }



\bibliography{sn2023xif,SN2014C}{}

\begin{thebibliography}{}
\expandafter\ifx\csname natexlab\endcsname\relax\def\natexlab#1{#1}\fi
\providecommand{\url}[1]{\href{#1}{#1}}
\providecommand{\dodoi}[1]{doi:~\href{http://doi.org/#1}{\nolinkurl{#1}}}
\providecommand{\doeprint}[1]{\href{http://ascl.net/#1}{\nolinkurl{http://ascl.net/#1}}}
\providecommand{\doarXiv}[1]{\href{https://arxiv.org/abs/#1}{\nolinkurl{https://arxiv.org/abs/#1}}}

\bibitem[{{Anders} \& {Grevesse}(1989)}]{1989GeCoA..53..197A}
{Anders}, E., \& {Grevesse}, N. 1989, \gca, 53, 197,
  \dodoi{10.1016/0016-7037(89)90286-X}

\bibitem[{{Astropy Collaboration} {et~al.}(2013){Astropy Collaboration},
  {Robitaille}, {Tollerud}, {Greenfield}, {Droettboom}, {Bray}, {Aldcroft},
  {Davis}, {Ginsburg}, {Price-Whelan}, {Kerzendorf}, {Conley}, {Crighton},
  {Barbary}, {Muna}, {Ferguson}, {Grollier}, {Parikh}, {Nair}, {Unther},
  {Deil}, {Woillez}, {Conseil}, {Kramer}, {Turner}, {Singer}, {Fox}, {Weaver},
  {Zabalza}, {Edwards}, {Azalee Bostroem}, {Burke}, {Casey}, {Crawford},
  {Dencheva}, {Ely}, {Jenness}, {Labrie}, {Lim}, {Pierfederici}, {Pontzen},
  {Ptak}, {Refsdal}, {Servillat}, \& {Streicher}}]{2013A&A...558A..33A}
{Astropy Collaboration}, {Robitaille}, T.~P., {Tollerud}, E.~J., {et~al.} 2013,
  \aap, 558, A33, \dodoi{10.1051/0004-6361/201322068}

\bibitem[{{Astropy Collaboration} {et~al.}(2018){Astropy Collaboration},
  {Price-Whelan}, {Sip{\H{o}}cz}, {G{\"u}nther}, {Lim}, {Crawford}, {Conseil},
  {Shupe}, {Craig}, {Dencheva}, {Ginsburg}, {VanderPlas}, {Bradley},
  {P{\'e}rez-Su{\'a}rez}, {de Val-Borro}, {Aldcroft}, {Cruz}, {Robitaille},
  {Tollerud}, {Ardelean}, {Babej}, {Bach}, {Bachetti}, {Bakanov}, {Bamford},
  {Barentsen}, {Barmby}, {Baumbach}, {Berry}, {Biscani}, {Boquien}, {Bostroem},
  {Bouma}, {Brammer}, {Bray}, {Breytenbach}, {Buddelmeijer}, {Burke},
  {Calderone}, {Cano Rodr{\'\i}guez}, {Cara}, {Cardoso}, {Cheedella}, {Copin},
  {Corrales}, {Crichton}, {D'Avella}, {Deil}, {Depagne}, {Dietrich}, {Donath},
  {Droettboom}, {Earl}, {Erben}, {Fabbro}, {Ferreira}, {Finethy}, {Fox},
  {Garrison}, {Gibbons}, {Goldstein}, {Gommers}, {Greco}, {Greenfield},
  {Groener}, {Grollier}, {Hagen}, {Hirst}, {Homeier}, {Horton}, {Hosseinzadeh},
  {Hu}, {Hunkeler}, {Ivezi{\'c}}, {Jain}, {Jenness}, {Kanarek}, {Kendrew},
  {Kern}, {Kerzendorf}, {Khvalko}, {King}, {Kirkby}, {Kulkarni}, {Kumar},
  {Lee}, {Lenz}, {Littlefair}, {Ma}, {Macleod}, {Mastropietro}, {McCully},
  {Montagnac}, {Morris}, {Mueller}, {Mumford}, {Muna}, {Murphy}, {Nelson},
  {Nguyen}, {Ninan}, {N{\"o}the}, {Ogaz}, {Oh}, {Parejko}, {Parley}, {Pascual},
  {Patil}, {Patil}, {Plunkett}, {Prochaska}, {Rastogi}, {Reddy Janga},
  {Sabater}, {Sakurikar}, {Seifert}, {Sherbert}, {Sherwood-Taylor}, {Shih},
  {Sick}, {Silbiger}, {Singanamalla}, {Singer}, {Sladen}, {Sooley},
  {Sornarajah}, {Streicher}, {Teuben}, {Thomas}, {Tremblay}, {Turner},
  {Terr{\'o}n}, {van Kerkwijk}, {de la Vega}, {Watkins}, {Weaver}, {Whitmore},
  {Woillez}, {Zabalza}, \& {Astropy Contributors}}]{2018AJ....156..123A}
{Astropy Collaboration}, {Price-Whelan}, A.~M., {Sip{\H{o}}cz}, B.~M., {et~al.}
  2018, \aj, 156, 123, \dodoi{10.3847/1538-3881/aabc4f}

\bibitem[{{Astropy Collaboration} {et~al.}(2022){Astropy Collaboration},
  {Price-Whelan}, {Lim}, {Earl}, {Starkman}, {Bradley}, {Shupe}, {Patil},
  {Corrales}, {Brasseur}, {N{\"o}the}, {Donath}, {Tollerud}, {Morris},
  {Ginsburg}, {Vaher}, {Weaver}, {Tocknell}, {Jamieson}, {van Kerkwijk},
  {Robitaille}, {Merry}, {Bachetti}, {G{\"u}nther}, {Aldcroft},
  {Alvarado-Montes}, {Archibald}, {B{\'o}di}, {Bapat}, {Barentsen},
  {Baz{\'a}n}, {Biswas}, {Boquien}, {Burke}, {Cara}, {Cara}, {Conroy},
  {Conseil}, {Craig}, {Cross}, {Cruz}, {D'Eugenio}, {Dencheva}, {Devillepoix},
  {Dietrich}, {Eigenbrot}, {Erben}, {Ferreira}, {Foreman-Mackey}, {Fox},
  {Freij}, {Garg}, {Geda}, {Glattly}, {Gondhalekar}, {Gordon}, {Grant},
  {Greenfield}, {Groener}, {Guest}, {Gurovich}, {Handberg}, {Hart},
  {Hatfield-Dodds}, {Homeier}, {Hosseinzadeh}, {Jenness}, {Jones}, {Joseph},
  {Kalmbach}, {Karamehmetoglu}, {Ka{\l}uszy{\'n}ski}, {Kelley}, {Kern},
  {Kerzendorf}, {Koch}, {Kulumani}, {Lee}, {Ly}, {Ma}, {MacBride}, {Maljaars},
  {Muna}, {Murphy}, {Norman}, {O'Steen}, {Oman}, {Pacifici}, {Pascual},
  {Pascual-Granado}, {Patil}, {Perren}, {Pickering}, {Rastogi}, {Roulston},
  {Ryan}, {Rykoff}, {Sabater}, {Sakurikar}, {Salgado}, {Sanghi}, {Saunders},
  {Savchenko}, {Schwardt}, {Seifert-Eckert}, {Shih}, {Jain}, {Shukla}, {Sick},
  {Simpson}, {Singanamalla}, {Singer}, {Singhal}, {Sinha}, {Sip{\H{o}}cz},
  {Spitler}, {Stansby}, {Streicher}, {{\v{S}}umak}, {Swinbank}, {Taranu},
  {Tewary}, {Tremblay}, {de Val-Borro}, {Van Kooten}, {Vasovi{\'c}}, {Verma},
  {de Miranda Cardoso}, {Williams}, {Wilson}, {Winkel}, {Wood-Vasey}, {Xue},
  {Yoachim}, {Zhang}, {Zonca}, \& {Astropy Project
  Contributors}}]{2022ApJ...935..167A}
{Astropy Collaboration}, {Price-Whelan}, A.~M., {Lim}, P.~L., {et~al.} 2022,
  \apj, 935, 167, \dodoi{10.3847/1538-4357/ac7c74}

\bibitem[{{Berger} {et~al.}(2023){Berger}, {Keating}, {Alexander}, {Cendes},
  {Eftekhari}, {Gurwell}, {Hiramatsu}, {Ho}, {Laskar}, {Margutti}, {Rao}, \&
  {Williams}}]{Berger23}
{Berger}, E., {Keating}, G., {Alexander}, K., {et~al.} 2023, Transient Name
  Server AstroNote, 131, 1

\bibitem[{{Bostroem} {et~al.}(2019){Bostroem}, {Valenti}, {Horesh}, {Morozova},
  {Kuin}, {Wyatt}, {Jerkstrand}, {Sand}, {Lundquist}, {Smith}, {Sullivan},
  {Hosseinzadeh}, {Arcavi}, {Callis}, {Cartier}, {Gal-Yam}, {Galbany},
  {Guti{\'e}rrez}, {Howell}, {Inserra}, {Kankare}, {L{\'o}pez}, {McCully},
  {Pignata}, {Piro}, {Rodr{\'\i}guez}, {Smartt}, {Smith}, {Yaron}, \&
  {Young}}]{Bostroem19}
{Bostroem}, K.~A., {Valenti}, S., {Horesh}, A., {et~al.} 2019, \mnras, 485,
  5120, \dodoi{10.1093/mnras/stz570}

\bibitem[{{Brethauer} {et~al.}(2022{\natexlab{a}}){Brethauer}, {Margutti},
  {Milisavljevic}, {Bietenholz}, {Chornock}, {Coppejans}, {De Colle}, {Hajela},
  {Terreran}, {Vargas}, {DeMarchi}, {Harris}, {Jacobson-Gal{\'a}n}, {Kamble},
  {Patnaude}, \& {Stroh}}]{Brethauer22}
{Brethauer}, D., {Margutti}, R., {Milisavljevic}, D., {et~al.}
  2022{\natexlab{a}}, \apj, 939, 105, \dodoi{10.3847/1538-4357/ac8b14}

\bibitem[{{Brethauer} {et~al.}(2022{\natexlab{b}}){Brethauer}, {Margutti},
  {Milisavljevic}, {Bietenholz}, {Chornock}, {Coppejans}, {De Colle}, {Hajela},
  {Terreran}, {Vargas}, {DeMarchi}, {Harris}, {Jacobson-Gal{\'a}n}, {Kamble},
  {Patnaude}, \& {Stroh}}]{2022ApJ...939..105B}
---. 2022{\natexlab{b}}, \apj, 939, 105, \dodoi{10.3847/1538-4357/ac8b14}

\bibitem[{{Brightman} \& {Nandra}(2011)}]{brightman11}
{Brightman}, M., \& {Nandra}, K. 2011, \mnras, 413, 1206,
  \dodoi{10.1111/j.1365-2966.2011.18207.x}

\bibitem[{{Chandra} {et~al.}(2023){Chandra}, {Chevalier}, {Nayana}, {Maeda}, \&
  {Ray}}]{chandra23}
{Chandra}, P., {Chevalier}, R., {Nayana}, A.~J., {Maeda}, K., \& {Ray}, A.
  2023, The Astronomer's Telegram, 16052, 1

\bibitem[{{Chandra} {et~al.}(2015){Chandra}, {Chevalier}, {Chugai}, {Fransson},
  \& {Soderberg}}]{Chandra15}
{Chandra}, P., {Chevalier}, R.~A., {Chugai}, N., {Fransson}, C., \&
  {Soderberg}, A.~M. 2015, \apj, 810, 32, \dodoi{10.1088/0004-637X/810/1/32}

\bibitem[{{Chandra} {et~al.}(2012){Chandra}, {Chevalier}, {Irwin}, {Chugai},
  {Fransson}, \& {Soderberg}}]{2012ApJ...750L...2C}
{Chandra}, P., {Chevalier}, R.~A., {Irwin}, C.~M., {et~al.} 2012, \apjl, 750,
  L2, \dodoi{10.1088/2041-8205/750/1/L2}

\bibitem[{{Chevalier} \& {Fransson}(2006)}]{Chevalier06}
{Chevalier}, R.~A., \& {Fransson}, C. 2006, \apj, 651, 381,
  \dodoi{10.1086/507606}

\bibitem[{Chevalier \& Fransson(2006)}]{Chevalier_2006}
Chevalier, R.~A., \& Fransson, C. 2006, The Astrophysical Journal, 651, 381,
  \dodoi{10.1086/507606}

\bibitem[{Chevalier \& Fransson(2017)}]{Chevalier17}
---. 2017, Handbook of Supernovae, 875–937,
  \dodoi{10.1007/978-3-319-21846-5_34}

\bibitem[{{Dessart} \& {John Hillier}(2022)}]{Dessart22}
{Dessart}, L., \& {John Hillier}, D. 2022, \aap, 660, L9,
  \dodoi{10.1051/0004-6361/202243372}

\bibitem[{{Dwarkadas} {et~al.}(2010){Dwarkadas}, {Dewey}, \&
  {Bauer}}]{Dwarkadas10}
{Dwarkadas}, V.~V., {Dewey}, D., \& {Bauer}, F. 2010, \mnras, 407, 812,
  \dodoi{10.1111/j.1365-2966.2010.16966.x}

\bibitem[{Evans {et~al.}(2010)Evans, Primini, Glotfelty, Anderson, Bonaventura,
  Chen, Davis, Doe, Evans, Fabbiano, Galle, Gibbs, Grier, Hain, Hall, Harbo,
  He, Houck, Karovska, Kashyap, Lauer, McCollough, McDowell, Miller, Mitschang,
  Morgan, Mossman, Nichols, Nowak, Plummer, Refsdal, Rots, Siemiginowska,
  Sundheim, Tibbetts, Stone, Winkelman, \& Zografou}]{evans10}
Evans, I.~N., Primini, F.~A., Glotfelty, K.~J., {et~al.} 2010, The
  Astrophysical Journal Supplement Series, 189, 37,
  \dodoi{10.1088/0067-0049/189/1/37}

\bibitem[{{Evans} {et~al.}(2007){Evans}, {Beardmore}, {Page}, {Tyler},
  {Osborne}, {Goad}, {O'Brien}, {Vetere}, {Racusin}, {Morris}, {Burrows},
  {Capalbi}, {Perri}, {Gehrels}, \& {Romano}}]{2007A&A...469..379E}
{Evans}, P.~A., {Beardmore}, A.~P., {Page}, K.~L., {et~al.} 2007, \aap, 469,
  379, \dodoi{10.1051/0004-6361:20077530}

\bibitem[{{Evans} {et~al.}(2009){Evans}, {Beardmore}, {Page}, {Osborne},
  {O'Brien}, {Willingale}, {Starling}, {Burrows}, {Godet}, {Vetere}, {Racusin},
  {Goad}, {Wiersema}, {Angelini}, {Capalbi}, {Chincarini}, {Gehrels}, {Kennea},
  {Margutti}, {Morris}, {Mountford}, {Pagani}, {Perri}, {Romano}, \&
  {Tanvir}}]{2009MNRAS.397.1177E}
---. 2009, \mnras, 397, 1177, \dodoi{10.1111/j.1365-2966.2009.14913.x}

\bibitem[{{Filippenko}(1997)}]{Filippenko97}
{Filippenko}, A.~V. 1997, \araa, 35, 309,
  \dodoi{10.1146/annurev.astro.35.1.309}

\bibitem[{Foreman-Mackey {et~al.}(2013)Foreman-Mackey, Hogg, Lang, \&
  Goodman}]{foreman-mackey_emcee:_2013}
Foreman-Mackey, D., Hogg, D.~W., Lang, D., \& Goodman, J. 2013, Publications of
  the Astronomical Society of the Pacific, 125, 306, \dodoi{10.1086/670067}

\bibitem[{{Fransson} {et~al.}(1996{\natexlab{a}}){Fransson}, {Lundqvist}, \&
  {Chevalier}}]{Fransson96}
{Fransson}, C., {Lundqvist}, P., \& {Chevalier}, R.~A. 1996{\natexlab{a}},
  \apj, 461, 993, \dodoi{10.1086/177119}

\bibitem[{{Fransson} {et~al.}(1996{\natexlab{b}}){Fransson}, {Lundqvist}, \&
  {Chevalier}}]{1996ApJ...461..993F}
---. 1996{\natexlab{b}}, \apj, 461, 993, \dodoi{10.1086/177119}

\bibitem[{{Gehrels} {et~al.}(2004){Gehrels}, {Chincarini}, {Giommi}, {Mason},
  {Nousek}, {Wells}, {White}, {Barthelmy}, {Burrows}, {Cominsky}, {Hurley},
  {Marshall}, {M{\'e}sz{\'a}ros}, {Roming}, {Angelini}, {Barbier}, {Belloni},
  {Campana}, {Caraveo}, {Chester}, {Citterio}, {Cline}, {Cropper}, {Cummings},
  {Dean}, {Feigelson}, {Fenimore}, {Frail}, {Fruchter}, {Garmire}, {Gendreau},
  {Ghisellini}, {Greiner}, {Hill}, {Hunsberger}, {Krimm}, {Kulkarni}, {Kumar},
  {Lebrun}, {Lloyd-Ronning}, {Markwardt}, {Mattson}, {Mushotzky}, {Norris},
  {Osborne}, {Paczynski}, {Palmer}, {Park}, {Parsons}, {Paul}, {Rees},
  {Reynolds}, {Rhoads}, {Sasseen}, {Schaefer}, {Short}, {Smale}, {Smith},
  {Stella}, {Tagliaferri}, {Takahashi}, {Tashiro}, {Townsley}, {Tueller},
  {Turner}, {Vietri}, {Voges}, {Ward}, {Willingale}, {Zerbi}, \&
  {Zhang}}]{2004ApJ...611.1005G}
{Gehrels}, N., {Chincarini}, G., {Giommi}, P., {et~al.} 2004, \apj, 611, 1005,
  \dodoi{10.1086/422091}

\bibitem[{{Grefensetette} {et~al.}(2017){Grefensetette}, {Harrison}, \&
  {Brightman}}]{Grefensetette17}
{Grefensetette}, B., {Harrison}, F., \& {Brightman}, M. 2017, The Astronomer's
  Telegram, 10427, 1

\bibitem[{{Harrison} {et~al.}(2013){Harrison}, {Craig}, {Christensen},
  {Hailey}, {Zhang}, {Boggs}, {Stern}, {Cook}, {Forster}, {Giommi},
  {Grefenstette}, {Kim}, {Kitaguchi}, {Koglin}, {Madsen}, {Mao}, {Miyasaka},
  {Mori}, {Perri}, {Pivovaroff}, {Puccetti}, {Rana}, {Westergaard}, {Willis},
  {Zoglauer}, {An}, {Bachetti}, {Barri{\`e}re}, {Bellm}, {Bhalerao},
  {Brejnholt}, {Fuerst}, {Liebe}, {Markwardt}, {Nynka}, {Vogel}, {Walton},
  {Wik}, {Alexander}, {Cominsky}, {Hornschemeier}, {Hornstrup}, {Kaspi},
  {Madejski}, {Matt}, {Molendi}, {Smith}, {Tomsick}, {Ajello}, {Ballantyne},
  {Balokovi{\'c}}, {Barret}, {Bauer}, {Blandford}, {Brandt}, {Brenneman},
  {Chiang}, {Chakrabarty}, {Chenevez}, {Comastri}, {Dufour}, {Elvis}, {Fabian},
  {Farrah}, {Fryer}, {Gotthelf}, {Grindlay}, {Helfand}, {Krivonos}, {Meier},
  {Miller}, {Natalucci}, {Ogle}, {Ofek}, {Ptak}, {Reynolds}, {Rigby},
  {Tagliaferri}, {Thorsett}, {Treister}, \& {Urry}}]{2013ApJ...770..103H}
{Harrison}, F.~A., {Craig}, W.~W., {Christensen}, F.~E., {et~al.} 2013, \apj,
  770, 103, \dodoi{10.1088/0004-637X/770/2/103}

\bibitem[{{Itagaki}(2023)}]{2023TNSTR1158....1I}
{Itagaki}, K. 2023, Transient Name Server Discovery Report, 2023-1158, 1

\bibitem[{{Jacobson-Gal{\'a}n} {et~al.}(2022){Jacobson-Gal{\'a}n}, {Dessart},
  {Jones}, {Margutti}, {Coppejans}, {Dimitriadis}, {Foley}, {Kilpatrick},
  {Matthews}, {Rest}, {Terreran}, {Aleo}, {Auchettl}, {Blanchard}, {Coulter},
  {Davis}, {de Boer}, {DeMarchi}, {Drout}, {Earl}, {Gagliano}, {Gall},
  {Hjorth}, {Huber}, {Ibik}, {Milisavljevic}, {Pan}, {Rest}, {Ridden-Harper},
  {Rojas-Bravo}, {Siebert}, {Smith}, {Taggart}, {Tinyanont}, {Wang}, \&
  {Zenati}}]{Jacobson-Galan22}
{Jacobson-Gal{\'a}n}, W.~V., {Dessart}, L., {Jones}, D.~O., {et~al.} 2022,
  \apj, 924, 15, \dodoi{10.3847/1538-4357/ac3f3a}

\bibitem[{{Mao} {et~al.}(2023){Mao}, {Zhang}, {Cai}, {Chen}, {Chen}, {Gao},
  {Li}, {Lyu}, {Qin}, {Sun}, {Xu}, {Zhang}, {Zhang}, {Zhao}, {Zheng}, {Zhou},
  \& {Ye}}]{2023TNSAN.130....1M}
{Mao}, Y., {Zhang}, M., {Cai}, G., {et~al.} 2023, Transient Name Server
  AstroNote, 130, 1

\bibitem[{{Margutti} {et~al.}(2014){Margutti}, {Milisavljevic}, {Soderberg},
  {Chornock}, {Zauderer}, {Murase}, {Guidorzi}, {Sanders}, {Kuin}, {Fransson},
  {Levesque}, {Chandra}, {Berger}, {Bianco}, {Brown}, {Challis},
  {Chatzopoulos}, {Cheung}, {Choi}, {Chomiuk}, {Chugai}, {Contreras}, {Drout},
  {Fesen}, {Foley}, {Fong}, {Friedman}, {Gall}, {Gehrels}, {Hjorth}, {Hsiao},
  {Kirshner}, {Im}, {Leloudas}, {Lunnan}, {Marion}, {Martin}, {Morrell},
  {Neugent}, {Omodei}, {Phillips}, {Rest}, {Silverman}, {Strader},
  {Stritzinger}, {Szalai}, {Utterback}, {Vinko}, {Wheeler}, {Arnett},
  {Campana}, {Chevalier}, {Ginsburg}, {Kamble}, {Roming}, {Pritchard}, \&
  {Stringfellow}}]{Margutti14}
{Margutti}, R., {Milisavljevic}, D., {Soderberg}, A.~M., {et~al.} 2014, \apj,
  780, 21, \dodoi{10.1088/0004-637X/780/1/21}

\bibitem[{{Margutti} {et~al.}(2017){Margutti}, {Kamble}, {Milisavljevic},
  {Zapartas}, {de Mink}, {Drout}, {Chornock}, {Risaliti}, {Zauderer},
  {Bietenholz}, {Cantiello}, {Chakraborti}, {Chomiuk}, {Fong}, {Grefenstette},
  {Guidorzi}, {Kirshner}, {Parrent}, {Patnaude}, {Soderberg}, {Gehrels}, \&
  {Harrison}}]{Margutti2017}
{Margutti}, R., {Kamble}, A., {Milisavljevic}, D., {et~al.} 2017, \apj, 835,
  140, \dodoi{10.3847/1538-4357/835/2/140}

\bibitem[{{Margutti} {et~al.}(2019){Margutti}, {Metzger}, {Chornock}, {Vurm},
  {Roth}, {Grefenstette}, {Savchenko}, {Cartier}, {Steiner}, {Terreran},
  {Margalit}, {Migliori}, {Milisavljevic}, {Alexander}, {Bietenholz},
  {Blanchard}, {Bozzo}, {Brethauer}, {Chilingarian}, {Coppejans}, {Ducci},
  {Ferrigno}, {Fong}, {G{\"o}tz}, {Guidorzi}, {Hajela}, {Hurley}, {Kuulkers},
  {Laurent}, {Mereghetti}, {Nicholl}, {Patnaude}, {Ubertini}, {Banovetz},
  {Bartel}, {Berger}, {Coughlin}, {Eftekhari}, {Frederiks}, {Kozlova},
  {Laskar}, {Svinkin}, {Drout}, {MacFadyen}, \& {Paterson}}]{Margutti2019}
{Margutti}, R., {Metzger}, B.~D., {Chornock}, R., {et~al.} 2019, \apj, 872, 18,
  \dodoi{10.3847/1538-4357/aafa01}

\bibitem[{{Matthews} {et~al.}(2023){Matthews}, {Margutti}, {Alexander},
  {Bright}, {Cendes}, {Berger}, {Lasker}, {Drout}, \&
  {Milisavljevic}}]{matthews23}
{Matthews}, D., {Margutti}, R., {Alexander}, K.~D., {et~al.} 2023, The
  Astronomer's Telegram, 16056, 1

\bibitem[{{Morozova} {et~al.}(2020){Morozova}, {Piro}, {Fuller}, \& {Van
  Dyk}}]{Morozova20}
{Morozova}, V., {Piro}, A.~L., {Fuller}, J., \& {Van Dyk}, S.~D. 2020, \apjl,
  891, L32, \dodoi{10.3847/2041-8213/ab77c8}

\bibitem[{{Morozova} {et~al.}(2018){Morozova}, {Piro}, \&
  {Valenti}}]{Morozova18}
{Morozova}, V., {Piro}, A.~L., \& {Valenti}, S. 2018, \apj, 858, 15,
  \dodoi{10.3847/1538-4357/aab9a6}

\bibitem[{{Ofek} {et~al.}(2014){Ofek}, {Sullivan}, {Shaviv}, {Steinbok},
  {Arcavi}, {Gal-Yam}, {Tal}, {Kulkarni}, {Nugent}, {Ben-Ami}, {Kasliwal},
  {Cenko}, {Laher}, {Surace}, {Bloom}, {Filippenko}, {Silverman}, \&
  {Yaron}}]{Ofek14}
{Ofek}, E.~O., {Sullivan}, M., {Shaviv}, N.~J., {et~al.} 2014, \apj, 789, 104,
  \dodoi{10.1088/0004-637X/789/2/104}

\bibitem[{{Pastorello} {et~al.}(2007){Pastorello}, {Smartt}, {Mattila},
  {Eldridge}, {Young}, {Itagaki}, {Yamaoka}, {Navasardyan}, {Valenti}, {Patat},
  {Agnoletto}, {Augusteijn}, {Benetti}, {Cappellaro}, {Boles}, {Bonnet-Bidaud},
  {Botticella}, {Bufano}, {Cao}, {Deng}, {Dennefeld}, {Elias-Rosa},
  {Harutyunyan}, {Keenan}, {Iijima}, {Lorenzi}, {Mazzali}, {Meng}, {Nakano},
  {Nielsen}, {Smoker}, {Stanishev}, {Turatto}, {Xu}, \&
  {Zampieri}}]{Pastorello07}
{Pastorello}, A., {Smartt}, S.~J., {Mattila}, S., {et~al.} 2007, \nat, 447,
  829, \dodoi{10.1038/nature05825}

\bibitem[{{Pastorello} {et~al.}(2008){Pastorello}, {Mattila}, {Zampieri},
  {Della Valle}, {Smartt}, {Valenti}, {Agnoletto}, {Benetti}, {Benn}, {Branch},
  {Cappellaro}, {Dennefeld}, {Eldridge}, {Gal-Yam}, {Harutyunyan}, {Hunter},
  {Kjeldsen}, {Lipkin}, {Mazzali}, {Milne}, {Navasardyan}, {Ofek}, {Pian},
  {Shemmer}, {Spiro}, {Stathakis}, {Taubenberger}, {Turatto}, \&
  {Yamaoka}}]{Pastorello08}
{Pastorello}, A., {Mattila}, S., {Zampieri}, L., {et~al.} 2008, \mnras, 389,
  113, \dodoi{10.1111/j.1365-2966.2008.13602.x}

\bibitem[{{Pastorello} {et~al.}(2013){Pastorello}, {Cappellaro}, {Inserra},
  {Smartt}, {Pignata}, {Benetti}, {Valenti}, {Fraser}, {Tak{\'a}ts}, {Benitez},
  {Botticella}, {Brimacombe}, {Bufano}, {Cellier-Holzem}, {Costado}, {Cupani},
  {Curtis}, {Elias-Rosa}, {Ergon}, {Fynbo}, {Hambsch}, {Hamuy}, {Harutyunyan},
  {Ivarson}, {Kankare}, {Martin}, {Kotak}, {LaCluyze}, {Maguire}, {Mattila},
  {Maza}, {McCrum}, {Miluzio}, {Norgaard-Nielsen}, {Nysewander}, {Ochner},
  {Pan}, {Pumo}, {Reichart}, {Tan}, {Taubenberger}, {Tomasella}, {Turatto}, \&
  {Wright}}]{Pastorello13}
{Pastorello}, A., {Cappellaro}, E., {Inserra}, C., {et~al.} 2013, \apj, 767, 1,
  \dodoi{10.1088/0004-637X/767/1/1}

\bibitem[{{Pastorello} {et~al.}(2018){Pastorello}, {Kochanek}, {Fraser},
  {Dong}, {Elias-Rosa}, {Filippenko}, {Benetti}, {Cappellaro}, {Tomasella},
  {Drake}, {Harmanen}, {Reynolds}, {Shappee}, {Smartt}, {Chambers}, {Huber},
  {Smith}, {Stanek}, {Christensen}, {Denneau}, {Djorgovski}, {Flewelling},
  {Gall}, {Gal-Yam}, {Geier}, {Heinze}, {Holoien}, {Isern}, {Kangas},
  {Kankare}, {Koff}, {Llapasset}, {Lowe}, {Lundqvist}, {Magnier}, {Mattila},
  {Morales-Garoffolo}, {Mutel}, {Nicolas}, {Ochner}, {Ofek}, {Prosperi},
  {Rest}, {Sano}, {Stalder}, {Stritzinger}, {Taddia}, {Terreran}, {Tonry},
  {Wainscoat}, {Waters}, {Weiland}, {Willman}, {Young}, \&
  {Zheng}}]{Pastorello18}
{Pastorello}, A., {Kochanek}, C.~S., {Fraser}, M., {et~al.} 2018, \mnras, 474,
  197, \dodoi{10.1093/mnras/stx2668}

\bibitem[{{Perley} {et~al.}(2023){Perley}, {Gal-Yam}, {Irani}, \&
  {Zimmerman}}]{2023TNSAN.119....1P}
{Perley}, D.~A., {Gal-Yam}, A., {Irani}, I., \& {Zimmerman}, E. 2023, Transient
  Name Server AstroNote, 119, 1

\bibitem[{{Perley} {et~al.}(2022){Perley}, {Sollerman}, {Schulze}, {Yao},
  {Fremling}, {Gal-Yam}, {Ho}, {Yang}, {Kool}, {Irani}, {Yan}, {Andreoni},
  {Baade}, {Bellm}, {Brink}, {Chen}, {Cikota}, {Coughlin}, {Dahiwale},
  {Dekany}, {Duev}, {Filippenko}, {Hoeflich}, {Kasliwal}, {Kulkarni}, {Lunnan},
  {Masci}, {Maund}, {Medford}, {Riddle}, {Rosnet}, {Shupe}, {Strotjohann},
  {Tzanidakis}, \& {Zheng}}]{Perley22}
{Perley}, D.~A., {Sollerman}, J., {Schulze}, S., {et~al.} 2022, \apj, 927, 180,
  \dodoi{10.3847/1538-4357/ac478e}

\bibitem[{{Quataert} \& {Shiode}(2012)}]{Quataert12}
{Quataert}, E., \& {Shiode}, J. 2012, \mnras, 423, L92,
  \dodoi{10.1111/j.1745-3933.2012.01264.x}

\bibitem[{{Riess} {et~al.}(2022){Riess}, {Yuan}, {Macri}, {Scolnic}, {Brout},
  {Casertano}, {Jones}, {Murakami}, {Anand}, {Breuval}, {Brink}, {Filippenko},
  {Hoffmann}, {Jha}, {D'arcy Kenworthy}, {Mackenty}, {Stahl}, \&
  {Zheng}}]{Riess22}
{Riess}, A.~G., {Yuan}, W., {Macri}, L.~M., {et~al.} 2022, \apjl, 934, L7,
  \dodoi{10.3847/2041-8213/ac5c5b}

\bibitem[{{Schlegel}(1990)}]{Schlegel90}
{Schlegel}, E.~M. 1990, \mnras, 244, 269

\bibitem[{{Smith}(2014)}]{Smith14}
{Smith}, N. 2014, \araa, 52, 487, \dodoi{10.1146/annurev-astro-081913-040025}

\bibitem[{{Smith} \& {Arnett}(2014)}]{Smith14b}
{Smith}, N., \& {Arnett}, W.~D. 2014, \apj, 785, 82,
  \dodoi{10.1088/0004-637X/785/2/82}

\bibitem[{{Soderberg} {et~al.}(2006){Soderberg}, {Chevalier}, {Kulkarni}, \&
  {Frail}}]{Soderberg062003bg}
{Soderberg}, A.~M., {Chevalier}, R.~A., {Kulkarni}, S.~R., \& {Frail}, D.~A.
  2006, \apj, 651, 1005, \dodoi{10.1086/507571}

\bibitem[{{Stritzinger} {et~al.}(2023){Stritzinger}, {Valerin}, {Elias-Rosa},
  {Fraser}, {Galbany}, {Gutierrez}, {Kankare}, {Kotak}, {Moran}, {Lundqvist},
  {Matilainen}, {Reguitti}, {Reynolds}, {Salmaso}, \&
  {Shappee}}]{2023TNSAN.145....1S}
{Stritzinger}, M., {Valerin}, G., {Elias-Rosa}, N., {et~al.} 2023, Transient
  Name Server AstroNote, 145, 1

\bibitem[{{Stroh} {et~al.}(2021){Stroh}, {Terreran}, {Coppejans}, {Bright},
  {Margutti}, {Bietenholz}, {De Colle}, {DeMarchi}, {Duran}, {Milisavljevic},
  {Murase}, {Paterson}, \& {Williams}}]{Stroh21}
{Stroh}, M.~C., {Terreran}, G., {Coppejans}, D.~L., {et~al.} 2021, \apjl, 923,
  L24, \dodoi{10.3847/2041-8213/ac375e}

\bibitem[{{Strotjohann} {et~al.}(2021){Strotjohann}, {Ofek}, {Gal-Yam},
  {Bruch}, {Schulze}, {Shaviv}, {Sollerman}, {Filippenko}, {Yaron}, {Fremling},
  {Nordin}, {Kool}, {Perley}, {Ho}, {Yang}, {Yao}, {Soumagnac}, {Graham},
  {Barbarino}, {Tartaglia}, {De}, {Goldstein}, {Cook}, {Brink}, {Taggart},
  {Yan}, {Lunnan}, {Kasliwal}, {Kulkarni}, {Nugent}, {Masci}, {Rosnet},
  {Adams}, {Andreoni}, {Bagdasaryan}, {Bellm}, {Burdge}, {Duev}, {Dugas},
  {Frederick}, {Goldwasser}, {Hankins}, {Irani}, {Karambelkar}, {Kupfer},
  {Liang}, {Neill}, {Porter}, {Riddle}, {Sharma}, {Short}, {Taddia},
  {Tzanidakis}, {van Roestel}, {Walters}, \& {Zhuang}}]{Strotjohann21}
{Strotjohann}, N.~L., {Ofek}, E.~O., {Gal-Yam}, A., {et~al.} 2021, \apj, 907,
  99, \dodoi{10.3847/1538-4357/abd032}

\bibitem[{{Thomas} {et~al.}(2022){Thomas}, {Wheeler}, {Dwarkadas}, {Stockdale},
  {Vink{\'o}}, {Pooley}, {Xu}, {Zeimann}, \& {MacQueen}}]{Thomas22}
{Thomas}, B.~P., {Wheeler}, J.~C., {Dwarkadas}, V.~V., {et~al.} 2022, \apj,
  930, 57, \dodoi{10.3847/1538-4357/ac5fa6}

\bibitem[{{Yamanaka} {et~al.}(2023){Yamanaka}, {Fujii}, \&
  {Nagayama}}]{2023arXiv230600263Y}
{Yamanaka}, M., {Fujii}, M., \& {Nagayama}, T. 2023, arXiv e-prints,
  arXiv:2306.00263, \dodoi{10.48550/arXiv.2306.00263}

\end{thebibliography}
\bibliographystyle{aasjournal}



\end{document}